# Remembering what we like:
# Toward an agent-based model of Web traffic

Bruno Gonçalves[1]*  Mark R. Meiss[1,2]  José J. Ramasco[3]
Alessandro Flammini[1]  Filippo Menczer[1,3]

[1]School of Informatics, Indiana University, Bloomington, IN, USA
[2]Advanced Network Management Lab, Indiana University, Bloomington, IN, USA
[3]Complex Networks and Systems Lagrange Laboratory (CNLL), ISI Foundation, Turin, Italy

## ABSTRACT

Analysis of aggregate Web traffic has shown that PageRank is a poor model of how people actually navigate the Web. Using the empirical traffic patterns generated by a thousand users over the course of two months, we characterize the properties of Web traffic that cannot be reproduced by Markovian models, in which destinations are independent of past decisions. In particular, we show that the diversity of sites visited by individual users is smaller and more broadly distributed than predicted by the PageRank model; that link traffic is more broadly distributed than predicted; and that the time between consecutive visits to the same site by a user is less broadly distributed than predicted. To account for these discrepancies, we introduce a more realistic navigation model in which agents maintain individual lists of bookmarks that are used as teleportation targets. The model can also account for branching, a traffic property caused by browser features such as tabs and the back button. The model reproduces aggregate traffic patterns such as site popularity, while also generating more accurate predictions of diversity, link traffic, and return time distributions. This model for the first time allows us to capture the extreme heterogeneity of aggregate traffic measurements while explaining the more narrowly focused browsing patterns of individual users.

## Categories and Subject Descriptors

H.3.4 [**Information Storage and Retrieval**]: Systems and Software—*Information networks*; H.4.3 [**Information Systems Applications**]: Communications Applications—*Information browsers*; H.5.4 [**Information Interfaces and Presentation**]: Hypertext/ Hypermedia—*Navigation*

*Corresponding author. Email: bgoncalv@indiana.edu

## 1. INTRODUCTION

Despite its simplicity, PageRank [2] has been a remarkably robust model of human Web browsing as a random surfing activity. Such models of Web surfing allow us to study how people interact with the Web. As more people spend a large portion of their time online, their Web traces provide an increasingly informative window into human dynamics. Models of user browsing also have important practical applications, ranging from ranking search results [2] to guiding crawlers [4] to predicting advertising revenues [6].

The availability of large volumes of Web traffic data is having two important consequences. On one hand, it motivates the integration of popularity measurements into search ranking algorithms [10]. On the other hand, it enables systematic testing of PageRank's underlying navigation assumptions [12]. Traffic patterns aggregated across users have revealed that some key assumptions—uniform random walk and uniform random teleportation—are widely violated, making PageRank a poor predictor of traffic. Such results leave open the question of how to design a better Web navigation model. Here we expand on our previous empirical analysis [12, 11] by considering also *individual* traffic patterns [8]. Our results provide further evidence for the limits of Markovian models such as PageRank. They suggest the need for an *agent-based* model that carries state information and can account for both individual and aggregate traffic patterns observed in real-world data.

We conducted a field study that allowed us to collect *individual* Web traffic data from over a thousand users on the main campus of Indiana University. Analysis of this data leads to several contributions, summarized as follows:

- We find that the traffic through hyperlinks is exceedingly more broadly distributed compared to the prediction of the PageRank model.
- We show that the empirical diversity of sites visited by individual users, as measured by Shannon entropy, is both smaller and more broadly distributed than predicted by PageRank.
- We find that the distribution of times between consecutive visits to the same site by a user is narrower than expected. Together with the previous item, this shows that a typical user has both focused interests and recurrent habits. We argue that the diversity apparent in many aggregate measures of traffic [13, 12] is a consequence of this diversity of individual interests rather than the behavior of extremely eclectic users who visit a wide variety of Web sites.

- We introduce a novel agent-based navigation model, *BookRank*, in which bookmarks are managed and used as teleportation targets, and tabbed browsing is a natural feature of the exploration process. The model reconciles individual browsing behavior with aggregate Web navigation patterns. One key individual behavior is revealed in a rank-based choice among previously visited sites as restart pages for surfing. Surprisingly, the mechanism behind this choice matches quantitatively the one reported for the selection among search engine results [5, 7].
- Finally, we demonstrate that the novel ingredients of *BookRank* allow it to improve considerably on PageRank's predictions for several empirical observations of both aggregate and individual traffic.

## 2. BOOKRANK MODEL

The introduction of PageRank [2] and the subsequent development of Google marked a turning point in the history of the Web. For the first time, search results were ranked using a model of the way people navigate through Web pages. Other models have been proposed over the years [9, 14, 1], but limited availability of empirical data on Web navigation has prevented a systematic test of their strengths and weaknesses. An alleged limitation of PageRank and similar models lies in the lack of user memory. An agent jumps from one page to another without any record of where it has been before or intends to go. Because they cannot purposefully return to a page, random surfers do not form navigational habits. Empirical measures of navigation patterns tell us that this assumption does not correspond to real user behavior [12, 11, 8].

*BookRank*, the model we propose here, provides random surfers with memory through the paradigm of bookmarks: each agent maintains a list of pages ranked by the number of previous visits. According to the *BookRank* model, agents in a population navigate the Web (in parallel or sequentially) according to the following algorithm.

Initially, each agent randomly selects a starting site (node). Then, for each time step:

1. Unless previously visited, the current node is added to the bookmark list. The frequency of visits is recorded, and the list of bookmarks is kept ranked from most to least visited.

2. With probability $p_t$, the agent teleports (jumps) to a previously visited site (bookmark). The bookmark with rank $R$ is chosen with probability $P(R) \propto R^{-\beta}$.

3. Otherwise, with probability $1 - p_t$, the user navigates locally, following a link from the present node. There are two alternatives:

   (a) With probability $p_b$, the back button is used leading back to the previous site.
   (b) Otherwise, with probability $1 - p_b$, an outgoing link is clicked with uniform probability.

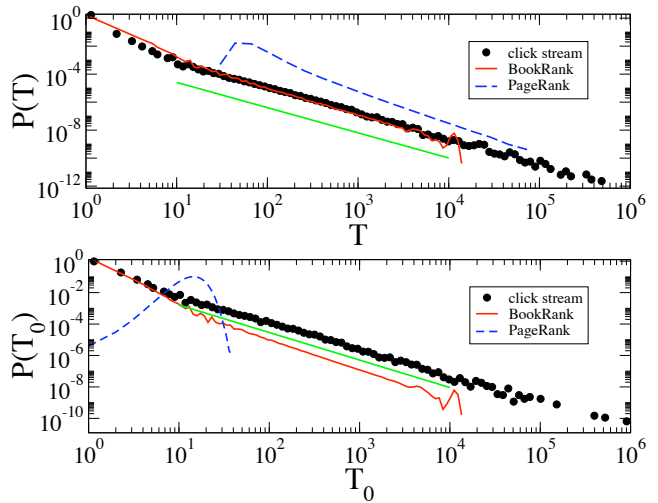

Figure 1: Distributions of all traffic to sites, $T$ (top) and traffic originating only from jumps to sites, $T_0$ (bottom) for the empirical click stream, PageRank and *BookRank*.

Note that PageRank is a special case of *BookRank* for $\beta = p_b = 0$.

## 3. EMPIRICAL DATA AND SIMULATIONS

To test the performance of our model, we compare its results (and those of PageRank) with empirical findings from a field study conducted on the main campus of Indiana University. The navigation data were gathered from a dedicated server located in the central routing facility of the Bloomington campus. This system had a 1 Gbps Ethernet port that received a mirror of all out bound network traffic from one of the undergraduate dormitories with more than 1000 recurrent users. HTTP request data was collected from this network feed over a period of about two months, from March 5 through May 3, 2008. A full analysis of the data set can be found elsewhere [11]. In our model we set $p_t = 0.15$, which matches the empirical data as well as the value traditionally used for PageRank [2].

In our simulations, agents navigate scale-free networks with $N$ nodes and degree distribution $P(k) \sim k^{-\gamma}$, generated according to the growth model of Fortunato *et al.* [15]. We set $N = 6.3 \times 10^5$ and $\gamma = 2.1$ to match the subset of the Web graph sampled in our data set. The functional form of $P(R)$ and the exponent $\beta \approx 1.4$ are also obtained by fitting the corresponding individual empirical data.

The simulation of both PageRank and *BookRank* models was repeated for a series of 250 different realizations of the network and performed using 1000 agents with $10^5$ clicks allocated to each.

### 3.1 Site Traffic

Let us first analyze the aggregate distribution of traffic received by sites. The traffic distribution $P(T)$ is displayed in Fig. 1 (top) for the two models along with our observed data. All distributions are well approximated by power laws $P(T) \sim T^{-\alpha}$. The exponent predicted by PageRank is $\alpha \approx 2.1$. Not surprisingly, this exponent matches that of the degree distribution in the Web graph [3]. This happens because random walkers visit the nodes of an undirected net-work with a probability proportional to their degree. The exponent changes very little when we introduce directed links and a small teleportation probability, as in PageRank [?]. The PageRank exponent does not reproduce well that of the empirical distribution, characterized by $\alpha \approx 1.75$ (see guide for the eye in Fig. 1). This discrepancy, together with the robustness of random walk exponents, indicates that the empirical data reflect some other mechanism beyond PageRank's uniform choice of the next destination. The *BookRank* model on the other hand provides an excellent prediction of the empirical $P(T)$.

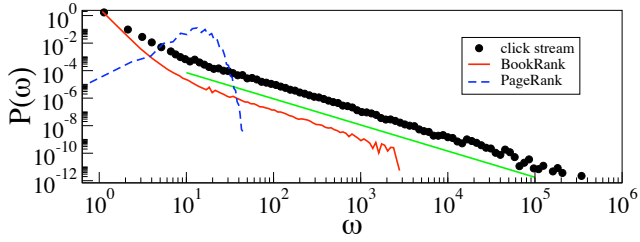

Figure 2: Distributions of link traffic $\omega$ for the empirical click stream, PageRank and *BookRank*.

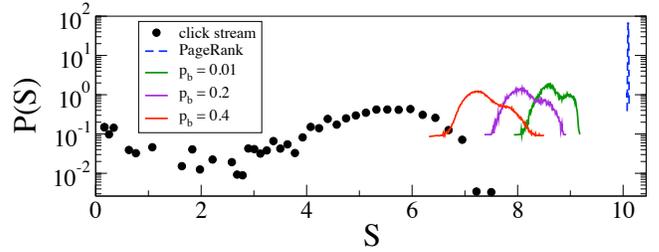

Figure 3: Distributions of Shannon entropy $S$ for the empirical click stream, PageRank and *BookRank*.

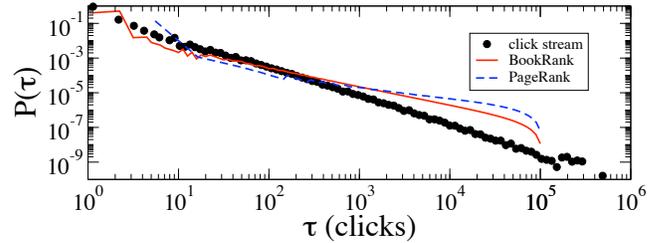

Figure 4: Distributions of time $\tau$ between consecutive visits to the same site for the empirical click stream, PageRank and *BookRank*.

Fig. 1 (bottom) shows the distribution of traffic originating from jumps, identified by HTTP requests with an empty referrer. Once again PageRank's prediction is poor; PageRank assumes a uniform probability $1/N$ of restarting from any node. In contrast, *BookRank* selects the next site using the bookmarks of each user; its prediction and the empirical data display a power-law behavior with close exponents $\alpha \approx 1.7$. Remarkably, following the same argument developed by Fortunato *et al.* [15], the exponent can be derived from $\beta$ via the relationship $\alpha = 1 + 1/\beta$. The fact that the empirical exponents $\alpha$ and $\beta$ satisfy this theoretical relation is a strong indication that our model's dependence on the rank of bookmarks for choosing teleportation targets quantitatively captures real user behavior. The exponent $\beta \approx 1.4$ is also surprisingly close to the one empirically observed in the distribution of click probability as a function of the ranks of results returned by a search engine [5, 7]. This provides further support for our hypothesis that rank-based choice is a key and general cognitive mechanism underlying Web search and browsing behavior.

### 3.2 Link Traffic

A second aggregate quantity of interest is the distribution of traffic across links, $P(\omega)$, as shown in Fig. 2. PageRank predicts a log-normal curve with well-defined mean and variance. The data instead reveal a much wider power-law distribution $P(\omega) \sim \omega^{-\delta}$ with $\delta \approx 1.9$ (see guide for the eye in Fig. 2). *BookRank* provides a much better approximation to the statistics of the link traffic than does PageRank.

### 3.3 Entropy distribution

One could easily be led to believe that the broad distributions characterizing aggregate user behavior are a reflection of the variability of traffic generated by single users, thus concluding that there is no such thing as a "typical" user from the point of view of traffic generated. The following analysis shows that this is not the case by better characterizing the navigation statistics of Web users.

To measure the diversity of the behavior of a group of users, we use Shannon information entropy $S = -\sum_i \rho_i \log \rho_i$ where $\rho_i$ is the fraction of visits of each particular user to site $i$. Given $n$ visits to a collection of sites, the entropy will be at its minimum $S = 0$ when all visits are to a single site, while its maximum $S = \log n$ is achieved when a single visit has been paid to each of $n$ distinct sites. Entropy offers a better probe into single user behavior than, say, the number of distinct sites visited; two users who have visited the same number of sites can have very different measures of entropy. It is therefore tempting to interpret the entropy as a measure of the information needed to describe the browsing pattern of a single user.

In Fig. 3 we compare the distribution $P(S)$ across users for the empirical data and the two models. PageRank yields a very narrow distribution of entropy, centered around high values, indicating a high degree of similarity among users' browsing behavior. The empirical data exhibits a wider distribution, centered around lower values. Given an equal number of visits, a real user will discover fewer pages than a PageRank surfer. The *BookRank* model predicts an entropy distribution closer to the empirical one, with a strong dependence on the parameter $p_b$ controlling the use of tabbed browsing (or the back button).

If we think of a user's browsing session as a tree [11], the quantity $p_b$ corresponds to the tree's branching ratio. Empirically this can be estimated by averaging across users. Since each user has multiple sessions, we can either macro-average the ratios (yielding an estimate $\langle p_b \rangle \approx 0.19 \pm 0.06$) or micro-average them (yielding a consistent but broader range $\langle p_b \rangle \approx 0.2 \pm 0.2$). In the simulations of Fig. 3 we used values of $p_b$ in this range; higher values shift the entropy distribution closer to the empirical one.

### 3.4 Times between visits

Another way to measure the spread of a user's activity through the network is to sample the time required to return to a page after a previous visit. The distribution of return times $\tau$, measured in number of clicks, is shown in Fig. 4.

Once again, the prediction generated by *BookRank* is significantly closer to the empirical $P(\tau)$. PageRank produces a distribution with a smaller power-law exponent, overestimating the time required for a user to return to a site. This is a consequence of its agents' lack of internal state. Our model, by introducing memory in the form of bookmarks, clearly improves on this result.

## 4. DISCUSSION

*BookRank* is a non-Markovian (history dependent) model that better approximates aggregate Web traffic than does PageRank. It introduces memory through the bookmarks maintained by each user, from which he or she restarts browsing after each jump. It also allows backtracking, which can account for the branching behavior characteristic of multi-tabbed browsing and use of the back button. We have shown that these two processes yield more realistic estimates of the number of sites visited by each user and the return time to previously visited pages, as compared with PageRank. Increasing the probability of branching, changing an otherwise linear browsing session into one involving multiple tabs, results in entropy values closer to the empirical ones. Furthermore, ranking bookmarks by the number of past visits is sufficient to reproduce the node and link traffic distributions. In summary, our model allows us to reproduce the intrinsic broadness of aggregate traffic measurements while better explaining the more narrowly focused browsing patterns of individual users.

Although this model is a clear step in the right direction, it shares some of the limitations present in previous models. The probability of ending a session is taken to be uniform, resulting in an exponential distribution of session lengths, while the empirical data obey a log-normal distribution [11]. Additionally the range of values for Shannon entropy does not completely overlap with our empirical observations. This may be addresses by enabling higher branching in the model; users may open many tabs from a page, which is hard to capture with just a back button.

In the current model, all users are stochastic replicas with identical parameters. An obvious way to extend the model is to take into account the diversity of the users with empirically driven distributions over their parameters (exponent $\beta$ of the probability distribution over bookmarks, jump probability $p_t$, back button usage $p_b$). Future work can also explore node-dependent jump probabilities to model the varying intrinsic relevance that users attribute to sites; for example, search engines are likely bookmarks for most users. Restrictions on the subset of nodes reachable by each user can be used to model different areas of interest. Finally, all pages are not equally likely to be the end point of a browsing session (i.e., the origin of a jump); our model can be extended to account for this heterogeneity.

### Acknowledgments

We thank the Advanced Network Management Laboratory at Indiana University and L. J. Camp of the IU School of Informatics for support and infrastructure. We also thank the network engineers of Indiana University for their support in deploying and managing the data collection system. This work was produced in part with support from the Institute for Information Infrastructure Protection research program. The I3P is managed by Dartmouth College and supported under Award 2003-TK-TX-0003 from the U.S. DHS, Science and Technology Directorate. BG was supported in part by NIH grant 1R21DA024259 and would like to thank A. Vespignani and S. Boettcher. This work is based upon work supported by NSF Award 0705676 and was supported in part by a gift to L. J. Camp from Google. Opinions, findings, conclusions, recommendations or points of view in this document are those of the authors and do not necessarily represent the official position of the DHS, I3P, NSF, Indiana University, Google, or Dartmouth College.